\documentclass[12pt]{article}
\usepackage{amsfonts,amssymb,amsmath}
\usepackage{epsfig}

\newcommand\RR{\mathbb R}
\newcommand\CC{\mathbb C}
\renewcommand{\Re}{\mathop{\mathrm{Re}}}
\renewcommand{\Im}{\mathop{\mathrm{Im}}}

\newcommand{\sign}{\mathop{\mathrm{sign}}}

\newcommand\beq{\begin{equation}}
\newcommand\eeq{\end{equation}}

\newtheorem{theorem}{Theorem}

\newtheorem{remark}{Remark}
\newtheorem{proposition}{Proposition}
\newtheorem{statement}{Statement}

\begin{document}

\title{Faddeev eigenfunctions for point potentials in two dimensions
\thanks{The work was supported by the  the Russian Federation Government grant
No~2010-220-01-077. The first author was also supported by  by LIA
``Physique th\'eorique et mati\`ere condens\'ee''  ENS - LANDAU and by the
program  ``Fundamental problems of nonlinear dynamics'' of the Presidium of
RAS.}}
\author{P.G. Grinevich
\thanks{Landau Institute of Theoretical Physics, Kosygin street 2,
117940 Moscow, Russia; e-mail: pgg@landau.ac.ru.} \and R.G.Novikov\thanks
{CNRS (UMR 7641), Centre de Math\'ematiques Appliqu\'ees,
\'Ecole Polytechnique, 91128, Palaiseau, France;
e-mail: novikov@cmap.polytechnique.fr}}
\date{}
\maketitle
\begin{abstract}
We present explicit formulas for the Faddeev eigenfunctions and related
generalized scattering data for point (delta-type) potentials in two
dimensions. In particular, we obtain the first explicit examples of
such eigenfunctions with contour singularity in spectral parameter at a
fixed real energy.
\end{abstract}

\section{Introduction}
Consider the two-dimensional Schr\"odinger equation
\beq
\label{eq:1}
-\Delta\psi + v(x)\psi = E\psi,  \ \ x\in\RR^2,
\eeq
where $v(x)$ is a real-valued sufficiently regular function on $\RR^2$ with
sufficient decay at infinity.

For (\ref{eq:1}) we consider the classical scattering eigenfunctions $\psi^+$
specified by
\beq
\label{eq:3}
\psi^+= e^{ikx}-i\pi\sqrt{2\pi}e^{-\frac{i\pi}{4}}f\left(k,|k|\frac{x}{|x|}
\right)
\frac{e^{i|k||x|}}{\sqrt{|k||x|}} +o\left(\frac{1}{\sqrt{|x|}} \right), \ \
\mbox{as} \ \ |x|\rightarrow\infty,
\eeq
$k\in\RR^2$, $k^2=E>0$, where a priori unknown function $f(k,l)$,
$k,l\in\RR^2$, $k^2=l^2=E$, arising in (\ref{eq:3}), is the classical
scattering amplitude for (\ref{eq:1}). In addition, we
consider the Faddeev eigenfunctions $\psi$ for (\ref{eq:1}) (see \cite{F1},
\cite{N}, \cite{G}), specified by
\beq
\label{eq:4}
\psi= e^{ikx}\left(1+o(1) \right) \ \
\mbox{as} \ \ |x|\rightarrow\infty,
\eeq
$k\in\CC^2$, $\Im k\ne 0$, $k^2=E$. The generalized scattering data
arise in more precise version of the expansion (\ref{eq:4}) (see also formulas
(\ref{eq:18})-(\ref{eq:21})). The Faddeev
eigenfunctions are quite important for inverse scattering (see, for example,
\cite{F2}, \cite{NKh}, \cite{G}).

In addition, as regards basic results on inverse scattering at fixed
energy in two dimensions see \cite{GM}, \cite{BLMP}, \cite{GN}, \cite{N},
\cite{N2}, \cite{G} and references therein. In addition,
as regards potentials, for which direct and inverse scattering at fixed energy
in two dimensions is exactly solvable see \cite{G}, \cite{TT} and references
therein.

However, modern monochromatic 2D inverse scattering is well-developed
under the assumption, that the Faddeev eigenfunctions has no singularities
in spectral parameter at fixed energy in complex domain. Due to $\cite{GN}$
this condition is restrictive. In particular, in \cite{GN} it was shown,
that at a negative energy  $E$ above the ground state one can expect
contour singularities in the complex plane of spectral parameter $\lambda$,
in typical situation. But the theory of generalized analytic functions,
used for monochromatic inverse scattering (see equations
(\ref{eq:26})-(\ref{eq:28}) on Faddeev eigenfunctions) is not developed for
such singularities.

In addition, no example of potential, for which direct and inverse
scattering in two dimensions is explicitly solvable for each energy
from some non-empty open interval, was given in literature. Besides, no example
of potential, for which the Faddeev eigenfunctions at a fixed energy in two
dimensions are calculated explicitly and have the aforementioned contour
singularities, was given in literature. May be the latter example for zero
energy can be extracted from \cite{TT} (private communication by
I.A. Taimanov).

In the present article we consider equation~(\ref{eq:1}), where $v(x)$ is
the 2-dimensional analog of the 3-dimensional point potential of
Zeldovich \cite{Z} and  Berezin-Faddeev \cite{BF}. Following
\cite{BF} we will write
\beq
\label{eq:2}
v(x)=\varepsilon\delta(x),
\eeq
but the precise sense of this potential will be specified below
(see Section~\ref{sect:3}), and strictly
speaking, $\delta(x)$ is not the standard Dirac delta-function. We show, that
this potential can be considered as an explicitly solvable model for direct
and inverse scattering for (\ref{eq:1}) at each real energy $E$.
In particular, we obtain explicit formulas for related Faddeev
eigenfunctions. These eigenfunctions have the aforementioned contour
singularities in spectral parameter $\lambda$ if $|E|$ is sufficiently small.
We hope, that this example will help to find correct formulation of
monochromatic inverse scattering in two dimensions in the presence of
spectral contour singularities.

Generalizations of the results of the present article for a sum of several
point potentials in 2D and in 3D will be given elsewhere.

\section{Some preliminaries}

It is convenient to write
\beq
\label{eq:5}
\psi= e^{ikx}\mu,
\eeq
where $\psi$ solves (\ref{eq:4}) and $\mu$ solves
\beq
\label{eq:6}
-\Delta\mu -2i k\nabla\mu+ v(x) \mu=0, \ \ \ \ k\in\CC^2, \ \
k^2=E.
\eeq

In addition, to relate eigenfunctions and scattering data it is
convenient to use the following presentations, used, for example,
in \cite{N2} for regular potentials:

\beq
\label{eq:18}
\mu^+(x,k)=1 -\int\limits_{\RR^2} \frac{e^{i\xi x}F(k,-\xi)}
{\xi^2+2(k+i0k)\xi } d\xi, \ \ k\in\RR^2\backslash0,
\eeq
\beq
\label{eq:18.1}
\mu_{\pm}(x,k)=1 -\int\limits_{\RR^2} \frac{e^{i\xi x}
H_{\pm}(k,-\xi)}{\xi^2+2(k\pm i0k_{\bot})\xi } d\xi, \ \ k\in\RR^2\backslash0,
\eeq
where $k_{\bot}=(-k_2,k_1)$ for $k=(k_1,k_2)$,
\beq
\label{eq:19}
\mu(x,k)=1 -\int\limits_{\RR^2} \frac{e^{i\xi x}H(k,-\xi)}{\xi^2+2k\xi } d\xi,
\ \ k\in\CC^2, \ \ \Im k\ne 0,
\eeq
where $\psi^+= e^{ikx}\mu^+$ are the eigenfunctions, specified by (\ref{eq:3}),
$\psi= e^{ikx}\mu$  are the eigenfunctions, specified by (\ref{eq:4}),
$\mu_{\pm}(x,k)=\mu(x,k\pm i 0 k_{\bot})$, $k\in\RR^2\backslash0$.

The following formulas holds:
\beq
\label{eq:20}
f(k,l)=F(k,k-l), \ \  k,l\in\RR^2, \ \ k^2=l^2=E>0,
\eeq
\beq
\label{eq:20.1}
h_{\pm}(k,l)=H_{\pm}(k,k-l), \ \  k,l\in\RR^2, \ \ k^2=l^2=E>0,
\eeq
\beq
\label{eq:21}
a(k)=H(k,0), \ \ b(k)=H(k,2\Re k), \ \  k\in\CC^2\backslash\RR^2, \ \
k^2=E\in\RR,
\eeq
where $f(k,l)$ is the classical scattering amplitude (\ref{eq:3}),
$h_{\pm}(k,l)$, $a(k)$, $b(k)$ are Faddeev generalized scattering data.

Let us recall, that the fixed energy surface $\Sigma_E=\{k\in\CC^2:k^2=E \}$
can be parametrized as follows:
\beq
\label{eq:24}
\Sigma_0=\Sigma_0^+\cup \Sigma_0^-, \ \
\Sigma_0^{\pm}=\{k_0^{\pm}(\lambda)=(\lambda,\pm i\lambda):\lambda\in\CC\},
\eeq
\beq
\label{eq:25}
\Sigma_E= \left\{k_E(\lambda)=\left(\left(\frac{1}{\lambda}
+\lambda \right)\frac{\sqrt{E}}{2}, \left(\frac{1}{\lambda}
-\lambda \right)\frac{i\sqrt{E}}{2}    \right)
:\lambda\in\CC \right\}, \ \ E\ne0.
\eeq
In addition:
\beq
\label{eq:25.1}
|\Re k_0^{\pm}(\lambda)|+ |\Im k_0^{\pm}(\lambda)| = 2 |\lambda|,
\eeq
\beq
\label{eq:25.1}
|\Re k_E(\lambda)|+ |\Im k_E(\lambda)| = \left\{\begin{aligned}
& \sqrt{|E|} |\lambda|, &  |\lambda|\ge 1,\\
& \sqrt{|E|} |\lambda|^{-1}, &  |\lambda|< 1,
 \end{aligned} \right. \ \ E\ne0.
\eeq

Note, that for regular real-valued potentials the following formulas hold
(at least outside the singularities of Faddeev functions in spectral parameter):
\beq
\label{eq:26}
\frac{\partial}{\partial\bar\lambda}\psi(x,k_E(\lambda)) =
\frac{\pi\sign(\lambda\bar\lambda-1)}{\bar\lambda} b(k_E(\lambda))
\overline{\psi(x,k_E(\lambda))}, \ \  E<0 \ \ \lambda\in\CC,
\eeq
\beq
\label{eq:27}
\frac{\partial}{\partial\bar\lambda}\psi(x,k_0^{\pm}(\lambda)) =
\frac{\pi}{\bar\lambda}b(k_0^{\pm}(\lambda))
\overline{\psi(x,k_0^{\pm}(\lambda))}, \ \ \lambda\in\CC\backslash0,
\eeq
\beq
\label{eq:28}
\frac{\partial}{\partial\bar\lambda}\psi(x,k_E(\lambda)) =
\frac{\pi\sign(\lambda\bar\lambda-1)}{\bar\lambda} b(k_E(\lambda))
\overline{\psi(x,k_E(\lambda))}, \  E>0, \  \lambda\in\CC, \ |\lambda|\ne1,
\eeq
\begin{align}
\label{eq:29}
&\psi_{\pm}(x,k_E(\lambda)) =\psi^+(x,k_E(\lambda))+\pi i
\int\limits_{|\lambda'|=1} h_{\pm}(k_E(\lambda), k_E(\lambda'))
\times
\nonumber\\
&\times\chi_+\left(\pm i\left(\frac{\lambda}{\lambda'}-
\frac{\lambda'}{\lambda} \right) \right) \psi^+(x,k_E(\lambda'))
|d\lambda'|, \ \ \ E>0, \ \lambda,\lambda'\in\CC,
\ |\lambda|=1,
\end{align}
\begin{align}
\label{eq:30}
&h_{\pm}(k_E(\lambda),k_E(\lambda')) -\pi i
\int\limits_{|\lambda''|=1} h_{\pm}(k_E(\lambda), k_E(\lambda''))
\chi_+\left(\pm i\left(\frac{\lambda}{\lambda''}-
\frac{\lambda''}{\lambda} \right) \right)
\times
\nonumber\\
&\times f(k_E(\lambda''),k_E(\lambda'))
|d\lambda''|=f(k_E(\lambda),k_E(\lambda')), \\
&\hspace{6cm} E>0, \ \lambda,\lambda',\lambda''\in\CC,
\ |\lambda|=  |\lambda'|  =1,\nonumber
\end{align}
\beq
\label{eq:31}
\mu(x,k_E(\lambda))\rightarrow1 \ \ \mbox{for} \ \ \lambda\rightarrow\infty, \ \
E\in\RR,
\eeq
where
$$
\chi_+(s)=1 \ \ \mbox{for} \ \ s>0, \ \  \chi_+(s)=0 \ \ \mbox{for} \ \ s\le0,
$$
and $\mu$ is related with $\psi$ as in (\ref{eq:5});
see \cite{GM}, \cite{BLMP}, \cite{GN}, \cite{N}, \cite{G} and references
therein. In  particular, formulas of the type (\ref{eq:29}),
(\ref{eq:30}) go back to \cite{F2}, formulas of the type (\ref{eq:26})-(\ref{eq:28})
go back to \cite{ABJF}, \cite{BC} .
In addition, let us recall that formulas (\ref{eq:26})-(\ref{eq:31}) give a basis
for monochromatic inverse scattering for regular real-valued potentials in two
dimensions.

\section{Main results}
\label{sect:3}
By analogy with \cite{BF} we understand the point-like potential
$v(x)$ from (\ref{eq:2}) as a limit for $N\rightarrow+\infty$ of non-local
potentials $V_N(x,x')=
\varepsilon(N) u_N(x) u_N(x') $ where,
\beq
\label{eq:7}
(V_N \mu)(x)= \varepsilon(N)\int\limits_{\RR^2} u_N(x) u_N(x') \mu(x') dx',
\eeq
\beq
\label{eq:8}
u_N(x) =\left(\frac{1}{2\pi}\right)^2 \int\limits_{\RR^2}
\hat u_N(\xi)e^{i\xi x}  d\xi, \ \
\hat u_N(\xi)=\left\{\small\begin{aligned} & 1 & |\xi|\le N, \\ &0 &   |\xi|> N,
\end{aligned}  \right.
\eeq
$ \varepsilon(N)$ is normalizing constant specified by (\ref{eq:13}).

For $v=V_N$ equation (\ref{eq:6}) has the following explicit solutions:
\beq
\label{eq:9}
\mu_N(x,k)= 1+ \left(\frac{1}{2\pi}\right)^2 \int\limits_{\RR^2}
\tilde\mu_N(\xi,k)e^{i\xi x}  d\xi,
\eeq
\beq
\label{eq:10}
\tilde\mu_N(\xi,k)=-\frac{\varepsilon(N) u_N(0)u_N(\xi)}
{1+ {\textstyle\varepsilon(N)}
 \left(\frac{1}{2\pi}\right)^2\int\limits_{\RR^2}
\frac{u_N(-\zeta)u_N(\zeta)}{\zeta^2+2k\zeta}  d\zeta} \cdot
\frac{1}{\xi^2+2k\xi},
\eeq
where $x\in\RR^2$,  $\xi\in\RR^2$,  $k\in\CC^2$, $\Im k\ne0$.
In addition, equation (\ref{eq:6}) has
the following classical scattering solutions:
\beq
\label{eq:11}
\mu^+_N(x,k)= \mu_N(x,k+i0k), \ \ x\in\RR^2, \ \ k\in\RR^2\backslash0,
\eeq
arising from
\beq
\label{eq:12}
\tilde\mu^+_N(\xi,k)=\tilde\mu_N(\xi,k+i0k), \ \
\xi\in\RR^2, \ \ k\in\RR^2\backslash0.
\eeq

\begin{theorem}
\label{th:1}
Let
\beq
\label{eq:13}
\varepsilon(N)=\frac{\alpha}{1-\frac{\alpha}{2\pi}\ln(N)}, \ \ \alpha\in\RR.
\eeq
Then:
\begin{enumerate}
\item The limiting eigenfunctions
\beq
\label{eq:13.1}
\psi(x,k)=e^{ikx} \lim\limits_{N\rightarrow+\infty} \mu_N(x,k), \ \
\psi^+(x,k)= e^{ikx} \lim\limits_{N\rightarrow+\infty} \mu^+_N(x,k)
\eeq
are well-defined for $x\in\RR^2$ and $k$ as indicated for (\ref{eq:3}),
(\ref{eq:4}).
\item  The following formulas hold:
\beq
\label{eq:14}
\psi(x,k)=e^{ikx}\left[
1 + \frac{\alpha}{1-\frac{\alpha} {2\pi}\ln(|\Re k| + |\Im k|) }\cdot g(x,k)\right],
\eeq
\beq
\label{eq:15}
 g(x,k)=-\left(\frac{1}{2\pi}\right)^2\int\limits_{\RR^2}
\frac{e^{i\xi x}}{\xi^2+2k\xi}  d\xi, \ \ k\in\CC^2, \ \ \Im k\ne 0, \ \
k^2=E\in\RR;
\eeq
\beq
\label{eq:16}
 \psi^+(x,k)=e^{ikx}\left[
1 + \frac{\alpha}{1+\frac{\alpha}
{4\pi}\left(\pi i - 2 \ln|k|\right)}\cdot g^+(x,k)\right],
\eeq
\beq
\label{eq:17}
 g^+(x,k)=-\left(\frac{1}{2\pi}\right)^2\int\limits_{\RR^2}
\frac{e^{i\xi x}}{\xi^2+2(k+i0k) \xi}  d\xi, \ \ k\in\RR^2, \ \ k\ne 0;
\eeq
\beq
\label{eq:16.1}
\psi_{\pm}(x,k)=\psi(x,k\pm i0k_{\bot} )= e^{ikx}\left[
1 + \frac{\alpha}{1-\frac{\alpha}
{2\pi}\ln|k|}\cdot g_{\pm}(x,k)\right],
\eeq
\beq
\label{eq:17.1}
 g_{\pm}(x,k)=-\left(\frac{1}{2\pi}\right)^2\int\limits_{\RR^2}
\frac{e^{i\xi x}}{\xi^2+2(k\pm i0k_{\bot}) \xi}  d\xi, \ \ k\in\RR^2, \ \ k\ne 0.
\eeq

\item The scattering data for the limiting potential
$v=\lim\limits_{N\rightarrow+\infty} V_N$, associated with the limiting
eigenfunctions $\psi^+(x,k)$,  $\psi_{\pm}(x,k)$,  $\psi(x,k)$  are given by:
\beq
\label{eq:22}
f(k,l)=\left(\frac{1}{2\pi}\right)^2 \frac{\alpha}{1+\frac{\alpha}
{4\pi}\left(\pi i - 2 \ln|k|\right)},
\eeq
\beq
\label{eq:22.1}
h_{\pm}(k,l) = \left(\frac{1}{2\pi}\right)^2  \frac{\alpha}{1-\frac{\alpha}
{2\pi}\ln(|k|) },
\eeq
where $k,l\in\RR^2$, $k^2=l^2=E>0$;
\beq
\label{eq:23}
a(k)=b(k) = \left(\frac{1}{2\pi}\right)^2  \frac{\alpha}{1-\frac{\alpha}
{2\pi}\ln(|\Re k| + |\Im k|) },
\eeq
where $k\in\CC^2$, $\Im k \ne 0$, $k^2=E\in\RR$.
\end{enumerate}
\end{theorem}

\begin{remark}
Relations between the absolute value of the scattering amplitude $f$ and its
phase for a two-dimensional point-like scatterer was given earlier
in \cite{BM}, see also \cite{BBMR} for further development. To our
knowledge no exact formulas for the Faddeev eigenfunctions $\psi$ and
related scattering data $a(k)$, $b(k)$ associated with 2D point potentials
were given in the literature.
\end{remark}

\begin{proposition}
\label{pr:1}
Formulas  (\ref{eq:26})-(\ref{eq:31}) are fulfilled for functions
$\psi=e^{ikx}\mu $, $\psi^+=e^{ikx} \mu^+$, $\psi_{\pm}=e^{ikx}\mu_{\pm}$,
$a$, $b$, $f$, $h_{\pm}$ of Theorem~\ref{th:1}, at least for $x\ne0$.
\end{proposition}
In addition, quite interesting properties of these functions for $\alpha\ne0$ can be
summarized as the following statement:
\begin{statement}
\label{st:1}
Let $\alpha\ne0$, $E_{1}=-\exp\left(\frac{4\pi}{\alpha}\right)$, $x\ne 0$.
Then:
\begin{enumerate}
\item If $E<E_{1}$, then the functions $\mu(x,k_E(\lambda))$,
$a(k_E(\lambda))$, $b(k_E(\lambda))$ are continuous in
$\lambda\in(\CC\cup\infty)$.
\item If $E=E_{1}$, then the functions $\mu(x,k_E(\lambda))$,
$a(k_E(\lambda))$, $b(k_E(\lambda))$  are singular in
$\lambda$ on the contour
$$
C_{E_1,\alpha}=T=\left\{\lambda\in\CC: |\lambda|=1 \right\}
$$
and are continuous in $\lambda\in(\CC\cup\infty)\backslash T $.
In addition, this energy level $E_1$ is discrete eigenvalue with
eigenfunction
\beq
\label{eq:32}
\psi_1(x)=-\left(\frac{1}{2\pi} \right)^2\int\limits_{\RR^2}
\frac{e^{i\xi x}}{\xi^2-E_1}d\xi.
\eeq
\item If $E_1<E<0$, then the functions $\mu(x,k_E(\lambda))$,
$a(k_E(\lambda))$, $b(k_E(\lambda))$ have simple singularities in
$\lambda$ on the contours
$$
C_{E,\alpha}=\left\{\lambda\in\CC: |\lambda|=\sqrt{\left|E/E_1\right| }\right\}
\cup \left\{\lambda\in\CC: |\lambda|=\sqrt{\left|E_1/E\right|}\right\}
$$
and are continuous in  $\lambda\in(\CC\cup\infty)\backslash C_{E,\alpha} $.
\item At zero energy $E=0$ the functions $\mu(x,k_0^{\pm}(\lambda))$,
$a(k_0^{\pm}(\lambda))$, $b(k_0^{\pm}(\lambda))$ have simple singularities in
$\lambda$ on the contour
$$
C_{0,\alpha}=\left\{\lambda\in\CC: |\lambda|=\frac{1}{2}
\sqrt{|E_1|}\right\}
$$
and are continuous in $\lambda\in\CC\backslash(C_{0,\alpha}\cup 0)$
on both components $\Sigma_0^+$,  $\Sigma_0^-$.
\item If $0<E<|E_1|$, then the functions $\mu(x,k_E(\lambda))$,
$a(k_E(\lambda))$, $b(k_E(\lambda))$ have simple singularities in
$\lambda$ on the contours
$$
C_{E,\alpha}=\left\{\lambda\in\CC: |\lambda|=\sqrt{\left|E/E_1\right| }\right\}
\cup \left\{\lambda\in\CC: |\lambda|=\sqrt{\left|E_1/E\right|}\right\}
$$
and are continuous in  $\lambda\in(\CC\cup\infty)\backslash (C_{E,\alpha}\cup
T)$.

In addition,
$\mu(x,k_E(\lambda(1\mp0))=\mu_{\pm}(x,k_E(\lambda))$ for $\lambda\in T$, where
$\mu_{\pm}$ are continuous in $\lambda\in T$ (whereas $a(k_E(\lambda))$,
$b(k_E(\lambda))$  are  continuous in a neighborhood of $T$).
\item If $E=|E_{1}|$, then the functions $\mu(x,k_E(\lambda))$,
$a(k_E(\lambda))$, $b(k_E(\lambda))$  are singular in
$\lambda$ on the contour
$$
T=\left\{\lambda\in\CC: |\lambda|=1 \right\},
$$
and are continuous in $\lambda\in(\CC\cup\infty)\backslash T $.
In addition, $\mu_{\pm}$, $h_{\pm}$ are not well-defined in this case;
for this energy we have real exceptional points for the Faddeev
eigenfunctions.
\item If $E>|E_1|$, then the functions $\mu(x,k_E(\lambda))$,
$a(k_E(\lambda))$, $b(k_E(\lambda))$ are continuous in
$\lambda\in(\CC\cup\infty)\backslash T$.

In addition,
$\mu(x,k_E(\lambda(1\mp0))=\mu_{\pm}(x,k_E(\lambda))$ for $\lambda\in T$, where
$\mu_{\pm}$ are continuous in $\lambda\in T$ (whereas $a(k_E(\lambda))$,
$b(k_E(\lambda))$  are  continuous in a neighborhood of $T$).
\end{enumerate}
\end{statement}

\section{Sketch of proofs}

The proof of items 1 and 2 of Theorem~\ref{th:1} follows from direct
calculations using the following formulas:
\begin{align}
\label{eq:33}
&\int\limits_{\xi\in\RR^2, \ |\xi|\le N} \frac{e^{i\xi x}}{\xi^2+2 k \xi-i0}
d\xi=
\int\limits_{\xi\in\RR^2, \ |\xi|\le N} \frac{e^{i\xi x}}{\xi^2+2 |k|\xi_1-i0}
d\xi= 2\pi \ln N + \nonumber \\
& +\pi^2i-4\int\limits_{0}^{\pi/2} \ln\left(2s\cos\phi\right) d\phi+
O(N^{-1}), \ \ k\in\RR^2\backslash0, \ \  N\rightarrow+\infty,
\end{align}
\begin{align}
\label{eq:34}
&\int\limits_{\xi\in\RR^2, \ |\xi|\le N} \frac{e^{i\xi x}}{\xi^2+2 k \xi}
d\xi =
\int\limits_{\xi\in\RR^2, \ |\xi|\le N} \frac{e^{i\xi x}}{\xi^2+2 |\Re k|\xi_1+
2i |\Im k|\xi_2}
d\xi= \nonumber \\
&  = 2\pi \ln N - 4\int\limits_{0}^{\pi/2} \ln\left(2\sqrt{ |\Re k|^2\cos^2\phi+
|\Im k|^2\sin^2\phi  } \right) d\phi+
O(N^{-1}), \\
& \hspace{5cm} k\in\CC^2\backslash\RR^2,  \ \ k^2=E\in\RR, \ \
N\rightarrow+\infty, \nonumber
\end{align}
where $\xi=(\xi_1,\xi_2)$;
\beq
\label{eq:35}
\int\limits_{0}^{\pi/2} \ln\left(\cos\phi \right) d\phi =-\frac{\pi}{2}\ln 2;
\eeq
\beq
\label{eq:36}
\int\limits_{0}^{\pi/2} \ln\left(a^2\cos^2\phi+
b^2\sin^2\phi  \right) d\phi=\pi\ln\frac{|a|+|b|}{2}, \ \ \ \
a,b\in\RR\backslash0.
\eeq
To prove item 3 of  Theorem~\ref{th:1} we use, in addition, formulas
(\ref{eq:18})-(\ref{eq:21}).

Proposition~\ref{pr:1} and Statement~\ref{st:1}
can be proved by direct calculations proceeding from Theorem~\ref{th:1} and
the well-known properties of the Faddeev Green function $G=e^{ikx}g$ (see, for
example, \cite{BLMP}, \cite{N}).


\begin{thebibliography}{ccc}

\bibitem{ABJF} M.J. Ablowitz, D. Bar Jaakov, A.S. Fokas, On the inverse
scattering of the time-dependent Schr\"odinger equation and the associated
Kadomtsev-Petviashvili equation, \textit{Stud. in applied math.} {\bf 69:2}
(1983), 135--143 .

\bibitem{BBMR} N.P. Badalyan, V.A. Burov, S.A.Morozov, O.D. Rumyantseva,
Scattering by acoustic boundary scattering with small wave sizes and their
reconstruction. \textit {Akusticheski\u{\i} Zhurnal}, {\bf 55:1} (2009), 3--10
(Rusian); English translation: \textit {Acoustical Physics} {\bf 55:1} (2009),
1--7.

\bibitem{BC} R.Beals, R.R.Coifman, Multidimensional inverse
scattering and nonlinear partial differential equations, \textit{Proc. of
Symposia in Pure Mathematics} {\bf 43} (1985) 45--70.

\bibitem{BF} F.A. Berezin and L.D. Faddeev, Remark on Schr\"odinger equation
with singular potential,
\textit{Dokl. Akad. Nauk SSSR} {\bf 137} (1961), 1011--1014 (Rusian);
English translation: \textit{Soviet Mathematics} {\bf 2} (1961), 372--375.

\bibitem{BLMP} M. Boiti, J. Leon, M. Manna, F. Pempinelli, On a
spectral transform of a KDV-like equation related to the Schr\"dinger
operator in the plane, \textit{Inverse Problems} {\bf 3}, 25--36 (1987).

\bibitem{BM} V.A. Burov, S.A.Morozov, Relationship between the amplitude and
phase of a signal scattering by a point-like acoustic inhomogeneity,
\textit {Akusticheski\u{\i} Zhurnal}, {\bf 47:6} (2001), 751--756
(Rusian); English translation: \textit {Acoustical Physics}, {\bf 47:6} (2001),
659--664.

\bibitem{F1} L.D. Faddeev, Growing solutions of the Schr\"odibger equation,
\textit{Dokl. Akad. Nauk SSSR} {\bf 165} (1965), 514--517 (Rusian);
English translation: \textit{Soviet Phys. Dokl.} {\bf 10} (1965), 1033--1035.

\bibitem{F2} L.D. Faddeev, Inverse problem of quantum scattering theory. II.
\textit{Itogi Nauki i Tekhniki. Sovr. Probl. Math. VINITI}, {\bf 3} (1974),
93--180 (Rusian); English translation:
\textit{Journal of Soviet Mathematics}, {\bf 5:3} (1976), 334--396.

\bibitem{G} P.G. Grinevich, The scattering transform for the two-dimensional
Schr\"odinger operator with a potential that decreases at infinity at
fixed nonzero energy, \textit{Uspekhi Mat. Nauk} {\bf 55:6(336)} (2000), 3--70
(Russian); English translation: \textit{Russian Math. Surveys} {\bf 55:6}
(2000), 1015--1083.

\bibitem{GM} P.G. Grinevich, S.V. Manakov, Inverse scattering
problem for the two-dimensional Schrodinger operator, $\bar\partial$ -
method and nonlinear equations, \textit {Funct. Anal. i ego Pril.}, {\bf 20:1}
(1986), 14--24 (Russian); English translation: \textit{Funct. Anal. Appl.},
{\bf 20}, (1986), 94--103.

\bibitem{GN} P.G. Grinevich, S.P. Novikov, Two-dimensional
'inverse scattering problem' for negative energies and generalized-analytic
functions. 1. Energies below the ground state,
\textit{Funct. Anal. i ego Pril.}{\bf 22:1} (1988), 23--33 (Rusian);
English translation: \textit{Funct. Anal. Appl.} {\bf 22} (1988), 19--27.

\bibitem{NKh} G.M. Henkin, R.G. Novikov,  The $\bar\partial$-equation in the
multidimensional inverse scattering problem, \textit{Uspekhi Mat. Nauk}
{\bf 42:3(255)} (1987), 93--152 (Russian); English translation:
\textit{Russian Math. Surveys} {\bf 42:3}  (1987) 109--180.

\bibitem{N} R.G. Novikov, The inverse scattering problem on a fixed energy
level for the two-dimensional Schr\"odinger operator. \textit{J. Funct. Anal.}
{\bf 103:2} (1992), 409--463.

\bibitem{N2} R.G. Novikov, Approximate solution of the inverse problem
of quantum scattering theory with fixed energy in dimension 2,
 \textit{Tr. Mat. Inst. Steklova} {\bf 225:2} (1999), Solitony Geom. Topol.
na Perekrest., 301--318 (Russian); English translation:
\textit{Proc. Steklov Inst. Math.} {\bf 225:2} (1999), 285--302

\bibitem{TT} I.A. Taimanov, S.P. Tsarev, On the Moutard transformation and
its applications to spectral theory and soliton equations. \textit{Sovrem.
Mat. Fundam. Napravl.} {\bf 35} (2010), 101--117,  (Russian).

\bibitem{Z}  Ya.B. Zel'dovich, Scattering by a singular potential in
perturbation theory and in the momentum representation. \textit{Z.
Eksper. Teoret. Fiz.} {\bf 38} (1960), 819--824 (Russian. English summary);
English translation: \textit{Soviet Physics. JETP} {\bf 11} (1960), 594--597.


\end{thebibliography}
\end{document}